\documentclass[12pt]{article}
\usepackage[dvips]{graphicx}
\usepackage{color}
\usepackage{amsfonts,amssymb}
\newcommand {\charex} {$K^+\mathrm{Xe} \rightarrow K^0 p \mathrm{Xe}'$}
\newcommand{\Tetat}{{\color{red}$\Theta^{+}$}$(p K_S^0)$({\color{red}1527$\pm$2.3})}
\newcommand{\Tetas}{{\color{red}$\Theta^+$}$(nK^+)$({\color{red}1555$\pm$10})}

\begin{document}

\vspace*{1cm}
\begin{center}
{\Large \bf What Can we Learn about $\Theta$ Baryon\\[1ex]
from Unified Picture for Hadron Spectra?}\\

\vspace{2mm}

{\large A.A. Arkhipov\footnote{e-mail: arkhipov@mx.ihep.su}\\
{\it State Research Center ``Institute for High Energy Physics" \\
 142281 Protvino, Moscow Region, Russia}}\\
\end{center}

\vspace{2mm}
\begin{abstract}
An analysis of the recent results from several experimental groups
reported observation of a new $\Theta$ baryon has been presented from
a view point of the unified picture for hadron spectra developed
early \cite{12}. It is shown that, in fact, two different $\Theta$
baryons have been discovered. We have also established that both
$\Theta$ baryons are excellently incorporated in the unified picture
for hadron spectra.  It is argued that the presented experimental
material  revealed an existence of the positive srangeness $\Theta$
partners for the observed $\Lambda$ and $\Sigma$ states with negative
strangeness as we predicted.
\end{abstract}

\section{Introduction}

The first experimental observation of a new, manifestly exotic baryon
(B=1, S=1, originally called the $Z^+$ but now denoted as the
$\Theta^+$) has been reported by LEPS Collaboration \cite{1}. The
experiment was carried out at the Laser-Electron Photon facility at
SPring-8 in Japan. A sharp baryon resonance peak for the strangeness
quantum number S=+1 was found at $1.54\pm 0.01$ GeV in the $K^-$
missing mass spectrum of the $\gamma n\rightarrow K^+K^-n$ reaction
on ${}^{12}C$. The width of the resonance was estimated to be smaller
than 25 MeV and Gaussian significance of the peak was 4.6 $\sigma$.
Soon after the confirmation of this observation was received by
several experimental groups \cite{2,3,4,5,6,7} from different
experiments where sharp peaks were observed in the $nK^+$ and
$pK_S^0$ invariant mass spectra at the mass near 1540 MeV. As a rule,
in all experiments a width was limited by the experimental
resolution.

DIANA Collaboration \cite{2} reported results from analysis of
bubble-chamber data for the charge-exchange reaction
$K^+Xe\rightarrow K^0pXe'$ where the spectrum of $K^0p$ effective
mass shows a resonant enhancement with M = 1.539$\pm$0.002 GeV and
$\Gamma\leq$ 9 MeV. The statistical significance of the enhancement
is near 4.4 $\sigma$.

A narrow peak in the $K^+n$ invariant mass spectrum that can be
attributed to exotic baryon with strangeness S=+1 was seen by CLAS
Collaboration \cite{3} in exclusive measurement of the reaction
$\gamma d\rightarrow K^+K^-pn$. This peak is at 1.542$\pm$0.005 GeV
with a measured width of 0.021 GeV, which is largely determined by
experimental mass resolution, and has a statistical significance 5.3
$\pm$0.5 $\sigma$.

The positive-strangeness baryon resonance $\Theta^+$ was observed in
photoproduction of the $nK^+K^0_s$ final state with the SAPHIR
detector at the Bonn Electron Stretcher Accelerator ELSA \cite{4}. It
was seen as a peak in the $nK^+$ invariant mass distribution with a
4.8$\sigma$ confidence level at the mass
$M_{\Theta^+}$=1540$\pm$4$\pm$2 MeV with an upper limit of the width
$\Gamma_{\Theta^+}\leq$25 MeV at 90\% c.l.

ITEP group \cite{5} reported an evidence for formation of a narrow
$K_S^0p$ resonance with the mass estimated as 1533$\pm$5 MeV in
neutrino and antineutrino collisions with nuclei. The observed width
is less than 20 MeV being entirely due to experimental resolution,
and statistical significance of the signal is near 6.7 standard
deviations. It is supposed that this resonance arises from neutrino
production of the $\Theta^+$ pentaquark baryon. The performed
analysis was based on the data obtained in past neutrino experiments
with big bubble chambers: WA21, WA25, WA59, E180, and E632.

A new, more comprehensive study of the $\Theta^+$ production on a
proton target has been done by CLAS Collaboration \cite{6}. Two
reactions, $\gamma p\rightarrow \pi^+K^+K^-n$ and $\gamma
p\rightarrow K^+K^-p$, were studied at Jefferson Lab using a tagged
photon beams with energy range of 3.5-4.7 GeV. A narrow baryon state
with strangeness S=+1 and mass M=1555$\pm$10 MeV was observed in the
$nK^+$ invariant mass spectrum. The peak's width is close to the
experimental mass resolution (FWHM =26 MeV) of the CLAS detector, and
its statistical significance is 7.8$\pm$1.0 $\sigma$. Besides, the
$pK^+$ invariant mass distribution was analyzed in the reaction
$\gamma p\rightarrow K^-K^+p$ too, and no resonance structure was
found there.

HERMES Collaboration \cite{7} presented the results of a search for
the $\Theta^+$ in quasi-real photoproduction on a deuterium target
through the decay channel $pK_S^0\rightarrow p\pi^+\pi^-$. A peak was
observed in the $pK_S^0$ invariant mass spectrum at 1528 $\pm$
2.6(stat) $\pm$ 2.1(syst) MeV. Depending on the background model, the
naive statistical significance of the peak is 4-6 standard deviations
and its width may be somewhat larger than the experimental resolution
of $\sigma$=4.3-6.2 MeV. In addition, no signal for the $\Theta^{++}$
baryon was observed in the $pK^+$ invariant distribution.

The most recent experimental evidence for the $\Theta^+$ came from
ZEUS Collaboration \cite{8} and SVD Collaboration \cite{9}. ZEUS
Collaboration presented preliminary results of an evidence for exotic
baryon decaying to $K_S^0$-(anti)proton: a signal at 1527$\pm$2(stat)
MeV with a Gaussian width of 10 $\pm$ 2 MeV and statistical
significance 4-5 $\sigma$ (from Gaussian fit) has been observed for
both $K^0$-proton and $K^0$-antiproton channels. SVD Collaboration
also reported an experimental observation a resonant structure with
M=1526 $\pm$ 3(stat) $\pm$ 3(syst) MeV and $\Gamma\leq$ 24 MeV in the
$pK_S^0$ invariant mass spectrum in the reaction $pA\rightarrow
pK_S^0 + X$.

It should be emphasized that the search for baryon resonances with
the strangeness S=+1, that cannot be built by three quarks, has a
long and interesting history, and first experimental observation of a
new, manifestly exotic baryon reported by LEPS Collaboration \cite{1}
was motivated in part by theoretical studies \cite{10} where baryon's
anti-decuplet has been constructed in the framework of chiral soliton
model exploiting the old and deep idea of Skyrme \cite{11}. Taking
the mass of $P_{11}$(1710) nucleon resonance as input identifying
that resonance with the member of the constructed anti-decuplet, it
was predicted a mass of $\sim$1530 MeV and a total width of less than
15 MeV for the $Z^+$ (spin 1/2, isospin 0 and strangeness +1) baryon
and pointed out that this region of the masses has avoided thorough
searches before. In particular, the S=+1 baryon resonances with a
relatively low mass of about 1530 MeV have not been searched for in
the $KN$ scattering in the past, probably, because momenta of kaons
were too high. Nevertheless, some claims for observing baryon states
decaying into $K^+n$ and $K^0p$ have been made in the past (see e.g.
old Particle Data Group Booklets in the middle of 70th) but they were
all with substantially higher mass and larger width than the
prediction in \cite{10}.

Since the simplest quark assignment consistent with the quantum
numbers of the $\Theta^+$ is ($uudd\bar s$), this baryon is often
called pentaquark. So, the first problem arises how to describe
pentaquark states in QCD. This problem is extremely complicated even
though on a pure empirical base e.g. what quark configuration is
favorable for the $\Theta^+$: is it diquark-triquark pentaquark state
or this is the state of antiquark and two diquarks or  something
else. It is well known that a description of new, recently discovered
charmonium meson states with unexpectedly low masses and narrow total
decay widths is extremely problematic in the framework of
phenomenology based on conventional quark potential models. Now, it
appears on a more high level with pentaquark states. Here, a similar
story repeats itself: the discovery of exotic baryon with low mass
and narrow width brings new, yet more serious problems for that
phenomenology. Concerning pure theory, up to now our theoretical
understanding of (especially low-energy) QCD is far from what is
desired. We don't know in quantum field theory how to construct even
the states for protons and pions though from fundamental quark and
gluon degrees of freedom in QCD, it appears to be very complicated
and unresolved problem so far. The best currently performed lattice
computations in QCD cannot help us to make calculations accurately at
low energy too. Nevertheless, there is a hope that powerful computers
will allow to overcome many significant technical problems in the
future.

In Ref. \cite{12}, where some of our previous studies were partially
summarized, it has been claimed that existence of the extra
dimensions in the spirit of Kaluza and Klein together with some novel
dynamical ideas may provide new conceptual issues for the global
solution of the spectral problem in hadron physics to create a
unified picture for hadron spectra. The question arises: what place
in the unified picture for hadron spectra the $\Theta$ baryon takes
up? This paper presents the answer to that question.

\section{Unified picture for hadron spectra: Understanding the
$\Theta$ baryon in comparison with the experiments}

Recently a new, very simple and at the same time quite general
physical law concerning the structure of hadron spectra has been
found; see \cite{12} and references therein. The developed
theoretical conception allowed to construct the global solution of
the spectral problem in hadron spectroscopy and provided quite new
scheme of systematics for hadron states. Our approach to hadron
spectroscopy has been verified with a large amount of experimental
data on meson's states and received an excellent agreement. What is
remarkable that all new meson states experimentally discovered last
year have been observed just at the masses predicted in our approach,
and those states appeared to be narrow as predicted too. The main
advantage of our developed theoretical conception is that all
calculated numbers for masses and widths do not depend on a special
dynamical model but follow from fundamental hypothesis on existence
of the extra dimensions with a compact internal extra space. One very
important fact has been established now in a reliable way: the size
of the internal compact extra space determines the global
characteristics of the hadron spectra while the masses of the
constituents are the fundamental parameters of the compound systems
being the elements of the global structure. Here we apply our
approach to analyse the hadron system under consideration i.e.
kaon-nucleon system.

First of all, according to the general law, the Kaluza-Klein tower of
KK-excitations for the kaon-nucleon system has been built by the
formula
\begin{equation}\label{kkexcit}
M_n^{NK} = \sqrt{m_N^2+\frac{n^2}{R^2}} +
\sqrt{m_K^2+\frac{n^2}{R^2}},\quad (n=1,2,3,...),
\end{equation}
where ($N=p,n$) and  ($K=K^0,K^+,K^-$),  $R$ is the fundamental scale
established before from analysis of nucleon-nucleon dynamics at low
energies; see \cite{12} and references therein for the details. The
such built Kaluza-Klein tower is shown in Table 1. Some experimental
information known from RPP Booklet has been presented in this Table
as well.

For completeness we have also calculated the Kaluza-Klein tower of
KK-excitations for the $\Lambda\pi$ system by similar formula like
(\ref{kkexcit}). That Kaluza-Klein tower is shown in Table 2 together
with experimental information extracted from the same RPP source. We
very hope that both Tables 1 and 2 might serve as a guide for the
experimenters in particle physics.

Now we  would like to compare the theoretically calculated spectra
with experimental material presented in above mentioned papers; see
Introduction. So then, let's see one after another.

\subsection{DIANA Collaboration experiment}

DIANA Collaboration investigated low-energy $K^+$Xe collisions by the
bubble chamber filled with liquid Xenon which has been exposed to a
separated $K^+$ beam with momentum of 850 MeV/c from the ITEP proton
synchrotron. The spectrum of $K^0 p$ effective mass was analyzed in
the charge-exchange reaction $K^+n \rightarrow K^0p$ where the
neutron was bound in a Xenon nucleus.  The events of this reaction
were fully measured and reconstructed in space using specially
designed stereo-projectors; the details on the experimental procedure
can be found in original paper \cite{2}.

Effective mass of the $K^0 p$ system formed in the charge-exchange
reaction is plotted in Fig. 1a for all measured events where an
enhancement is clear seen at the mass $M \simeq 1540$ MeV/c$^2$. To
estimate the level of background, the effective mass spectrum was
fitted to a linear combination of two regular distributions: one
distribution was obtained by simulation that took into account many
real conditions of the experiment, and a distribution which was
obtained by the method of random stars.  The result of the fit is
depicted by the dashed line in Fig. 1a. After that, additional
topological selections have been applied: see \cite{2} for details.
Of the 1112 measured events, nearly a half (541 events) survived the
additional selections. In the $K^0 p$ mass spectrum for these events,
that is shown in Fig. 1b, the enhancement near 1540 MeV/c$^2$ became
more prominent.

In summary it has been concluded that a baryon resonance with mass $M
= 1539\pm2$ MeV and width $\Gamma \le 9$ MeV has been observed in the
$K^0 p$ effective-mass spectrum. The statistical significance of the
signal was estimated as $4.4\sigma$. It is suggested that the
observed resonance is a strong indication for formation of the exotic
pentaquark $Z^+$ baryon, but the work is still in progress.

Vertical, spectral lines have been plotted in Figs. 1a and 1b as well
to compare the theoretically calculated spectrum of KK excitations in
the $K^0 p$ system with the picture observed at the experiment. Even
though we found a strong correlation of the spectral lines with the
peaks on the histograms the reported resonance peak appeared
approximately in the middle between two spectral lines corresponding
to $M_6^{pK^0}$(1527)-storey and $M_7^{pK^0}$(1558)-storey in KK
tower for $pK^0$ system; see Table 1. Let's assume that further,
thorough analysis would clarify a matter of dispute.

\subsection{CLAS Collaboration $\gamma d$ experiment}

CLAS Collaboration \cite{3} has performed an exclusive measurement on
deuterium for the reaction $\gamma d \rightarrow K^+ K^- p (n)$ where
the final state neutron was reconstructed from the missing momentum
and energy. The data have been obtained at the Thomas Jefferson
National Accelerator Facility with the CLAS detector and the photon
tagging system: photon beams were produced by 2.474 and 3.115 GeV
electrons incident on a bremsstrahlung radiator, the maximum tagged
photon energy was $95\%$ of the electron beam energy, the photons
struck a liquid-deuterium target; see \cite{3} and references therein
for details. The details on the experimental procedure as well as a
discussion of the explicit cuts which have been made to remove the
main background sources from the final event sample in order to
enhance the signal relative to background can be found in original
paper \cite{3} too.

Final result is presented as $nK^+$ invariant mass spectrum which is
shown in Fig. 2. A fit (solid line) to the peak and a Gaussian plus
constant term fit to the background (dashed line) are also depicted
in this Figure. The peak is at $1.542 \pm 0.005$ GeV with a width
(FWHM) of 0.021 GeV. The width is consistent with the instrumental
resolution. The uncertainty of 0.005 GeV in the mass is due to
calibration uncertainties of the photon tagging spectrometer, the
electron beam energy, and the momentum reconstruction in CLAS. The
statistical significance of the peak is $5.2 \pm 0.6\ \sigma$. The
spectrum of events removed by the $\Lambda$(1520) cut is shown in
Fig. 2 by the dashed-dotted histogram, and does not appear to be
associated with the peak at 1.542 GeV.

The invariant mass spectrum of $pK^+$ system was also examined by
CLAS Collaboration using the same event selection as in the case of
$nK^+$ system. The statistics were limited, but there was no clear
peak in the signal region. It was noted that the CLAS acceptance for
the $pK^+$ system is not the same as for the $nK^+$ system, so the
two spectra are not directly comparable. The absence of peak in
$M(pK^+)$ spectrum suggests that $\Theta^+$ is probably an
isosinglet, though there is no a firm conclusion based on the current
data.

We have plotted in Fig. 2 the spectral lines corresponding to KK
excitations in the $nK^+$ system taken from Table 1. Here it is also
found a strong correlation of the spectral lines with the peaks on
the histogram. The reported resonance peak appeared between two
spectral lines corresponding to $M_6^{nK^+}$(1525)-storey and
$M_7^{nK^+}$(1556)-storey in KK tower for $nK^+$ system though a
little bit closed in second spectral line of $M_7^{nK^+}$(1556).

\subsection{SAPHIR Collaboration experiment}

SAPHIR Collaboration \cite{4} presented evidence for the $\Theta^+$
in photoproduction of the $nK^+K^0_s$ final state off protons.
Actually, the reaction $\gamma p \to nK^0_sK^+$ with the decay
$K^0_s\to\pi^+\pi^-$ was measured with the SAPHIR detector at ELSA.
The ELSA electron beam produced photons via bremsstrahlung in a
copper foil radiator. The tagged photon energies were from 31\,\% to
94\,\% of the incident electron energy which was  2.8 GeV for the
data presented in this experiment. The photon beam passed through a
liquid hydrogen target. Non--interacting photons were detected in a
photon counter. The coincidences of tagger and photon counter
determined the photon flux. Additional, needed details concerning the
experimental procedure can be found in original paper \cite{4}.

The observation of the $\Theta^+$ in the reaction chain
\begin{equation}
\gamma p \to \Theta^{+} K^0_s , \qquad\ \Theta^{+}\to nK^+ , \qquad
K^0_s\to\pi^+\pi^- \label{eq:plus}
\end{equation}
has been presented by  Fig. 3 showing the $nK^+$ invariant mass
distribution after cuts; original paper \cite{4} contains a
comprehensive discussion of all cuts which have been made to arrive
at this Figure. As is seen from Fig. 3 there is a clear peak at $\sim
1540$\,MeV. The statistical significance of the peak is $5.2\sigma$.

The photoproduction of the $pK^+K^-$ final state has been studied by
SAPHIR Collaboration as well to search for the doubly charged
$\Theta^{++}$ in the reaction chain
\begin{equation}
\gamma p \to \Theta^{++} K^- , \qquad\ \Theta^{++}\to pK^+.
\label{eq:plusplus}
\end{equation}
Fig.~4 shows the resulting $pK^+$ invariant mass spectrum. It should
be emphasized that the data are consistent with a small structure in
the $pK^+$ invariant mass distribution. However, since the SAPHIR
acceptance was considerably larger for the fully constrained
$pK^+K^-$ events they expected a peak with more than 5000
$\Theta^{++}$, far above the observed level. The absence of a signal
in the $pK^+$ invariant mass distribution at the expected strength
resulted in conclusion that the $\Theta^+$ must be isoscalar.

As before we have plotted in Figs. 3 and 4 the spectral lines
corresponding to KK excitations in the $nK^+$ and $pK^+$ systems
taken from Table 1. Again we found a strong correlation of the
spectral lines with the peaks on the histograms. The reported
resonance peak of $\Theta^+$ in Fig.3 appears just in the middle
between two spectral lines corresponding to $M_6^{nK^+}$(1525)-storey
and $M_7^{nK^+}$(1556)-storey in KK tower for $nK^+$ system (see
Table 1), and this peak is broad enough. However, we would like to
point out a remarkable fact that the spectral line $M_7^{pK^+}$(1555)
in Fig.4 exactly coincided with the experimentally observed structure
in the $pK^+$ invariant mass distribution. In our opinion further,
scrupulous experimental studies of kaon-nucleon system are very
desirable, no doubt they would welcome.

\subsection{NEUTRINO interactions}

ITEP group reported a search for formation of the $\Theta^+$ baryon
in neutrino and antineutrino collisions with protons, deuterons, and
Neon nuclei \cite{5}. The data collected by several neutrino
experiments with big bubble chambers -- BEBC at CERN and the 15-foot
chamber at Fermilab -- have been analyzed. Compiled database included
about 120 000 $\nu_\mu$ and $\bar{\nu}_\mu$ induced charged-current
events, and covered the bulk of neutrino data collected with BEBC
(WA21, WA25, and WA59) and a significant fraction of data collected
with the 15-foot bubble chamber at Fermilab (E180 and E632).
Neutral-current interactions were not systematically included in the
database, and therefore the analysis was restricted to
charged-current events only. It has been stressed that neutrino data
from big bubble chambers are still unrivaled in quality and
completeness of physics information. Further details on the neutrino
experiments can be found in original paper \cite{5} and references
therein.

Figure 5 shows the $K_S^0p$ invariant mass distribution for the Neon
and Deuterium data combined (top panel). In the bottom panel the same
$K_S^0p$ distribution with bins shifted by 5 MeV has been plotted. A
fit of the bottom histogram yielded $M = 1533 \pm 5$ MeV and $\sigma
= 8.4 \pm 2.0$ MeV for the position and r.m.s. width of the
resonance. The statistical significance of the peak appeared to be
near 6.7 standard deviations.

We have plotted in Fig. 5 (top panel) the spectral lines
corresponding to KK excitations in the $pK^0$ system taken from Table
1 and found a striking picture: all spectral lines without any
exception coincided with the peaks on the histogram. The reported
resonance peak of $\Theta^+$ just corresponds to the spectral line
$M_6^{pK^0}$(1527) in KK tower for $pK^0$ system (see Table 1).
Really, analyzing the excellent neutrino data resulted in excellent
agreement with the theory.

\subsection{CLAS Collaboration $\gamma p$ experiment}

Recently a more comprehensive study of the $\Theta^+$ production on a
proton target which includes data from three distinct runs under
different experimental conditions in CLAS has been reported by CLAS
Collaboration \cite{6}. Two reactions, $\gamma p \rightarrow \pi^+
K^+K^- n$ and $\gamma p \rightarrow K^+K^- p$, have been analyzed in
the three runs having identical geometrical acceptance and trigger
requirement but with a slightly different tagged photon beams in the
energy range of 3.2--3.95~GeV, 3--5.25~GeV, and 4.8--5.47~ GeV,
respectively (details described in \cite{6}). The combined analysis
of these three runs offered access to a wider range of acceptance and
energies. The final $n K^+$ invariant mass spectrum calculated from
missing mass in the reaction $\gamma p\rightarrow \pi^+ K^- X$,
combining data from all three data runs, is shown  in  Fig.~6. The
authors concluded that no obvious structure is seen in this spectrum.

Figure 7 shows the resulting $nK^+$ mass spectrum prepared by
applying a several explicit angular cuts. As is seen the $\Theta^+$
peak was clearly observed in that Figure. The $nK^+$ effective mass
distribution shown in Fig. 7 was fitted  by the sum of a Gaussian
function and a background function obtained from the simulation. The
fit parameters are: $N_{\Theta^+}=41\pm10$, $M=1555\pm 1$ MeV, and
$\Gamma=26\pm7$ MeV (FWHM), where the errors are statistical. The
systematic mass scale uncertainty was estimated to be $\pm 10$ MeV.
It was also estimated the significance to be 7.8 $\pm$ 1.0~$\sigma$.

In addition, a search for a manifestly exotic baryon ($Q=2$, $S=+1$)
was performed in the reaction $\gamma p\rightarrow
K^-X^{++},~X^{++}\rightarrow pK^+$. Using the similar procedures as
in previous case, the $pK^+$ invariant mass distribution was built
but there were no resonance structures evident in that distribution.
However, a more detailed analysis will be presented in a future.

As usual, we have plotted in Figs. 6 and 7 the spectral lines
corresponding to KK excitations in the $nK^+$ system taken from Table
1. Here, in addition to strong correlation of the spectral lines with
the peaks on the histograms as in previous cases, we found out new,
quite a remarkable fact: The reported resonance peak of $\Theta^+$
just corresponds to the spectral line $M_7^{nK^+}$(1556) in KK tower
for the $nK^+$ system (see Table 1).

\subsection{HERMES Collaboration experiment}

The results of a search for the $\Theta^+$ in quasi-real
photoproduction  on deuterium have been presented by HERMES
Collaboration \cite{7}.  The data were obtained by the HERMES
experiment with the 27.6\,GeV positron beam of the HERA storage ring
at DESY. The analysis searched for inclusive photoproduction of the
$\Theta^+$ followed by the decay chain $\Theta^+\to p K^0_S \to p
\pi^+ \pi^-$. Events selected contained at least three tracks: two
oppositely charged pions in coincidence with one proton. The event
selection included constraints on the event topology to maximize the
yield of the $K^0_S$ peak in the $M_{\pi^+\pi^-}$ spectrum while
minimizing its background. However, no constraints were optimized to
increase the significance of the signal visible in the final $M_{p
\pi^+\pi^-}$ spectrum, as such optimization would have produced a
spectrum to which standard statistical tests do not apply. To search
for the $\Theta^+$, events were selected with a $M_{\pi^+\pi^-}$
invariant mass within $\pm 2\,\sigma$ about the centroid of the
$K_S^0$ peak. More details on the experimental procedure described in
original paper \cite{7} where the interested reader referred to. The
resulting spectrum of the invariant mass of the $p\pi^+\pi^-$ system
is displayed in Fig.~8. As is seen a narrow peak is observed, and
this peak is identified with the $\Theta^+$.

Figure 9 shows the spectra of invariant mass $M_{pK^-}$ (top) and
$M_{pK^+}$ (bottom). The event selection for the spectra in this
figure was the same as for the $p K^0_S$ analysis, except the
reconstructed $K^0_S$ track was replaced by that of the observed
charged kaon. The $\Lambda(1520)$ mass peak fitted with a Gaussian is
shown in the $M_{pK^-}$ spectrum of Fig.~9 (details of the fit can be
found in the original paper). A clear peak is seen for the
$\Lambda(1520)$ in the $M_{pK^-}$ invariant mass distribution.
However, no peak structure is seen for the hypothetical $\Theta^{++}$
in the $M_{pK^+}$ invariant mass distribution near 1.53\,GeV.

Following our general strategy, as before, we have plotted in Figs. 8
and 9 the spectral lines corresponding to KK excitations in the
$pK^0$ system (Fig. 8) and $pK^\pm$ system (Fig. 9) taken from Table
1. No doubt, this is quite a remarkable fact that the reported
resonance peak of $\Theta^+$ exactly coincided with the spectral line
$M_6^{pK^0}$(1527) in KK tower for the $pK^0$ system, and
$\Lambda(1520)$ peak just corresponds to the spectral line
$M_6^{pK^\pm}$(1523.6) in KK tower for the $pK^\pm$ system (see Table
1). Moreover, we observe a barely visible structures in the
$M_{pK^+}$ invariant mass distribution around the spectral lines
depicted in Fig. 9. Clearly, a much more careful experimental studies
with a higher statistics and resolution are utterly desired.

\subsection{SVD Collaboration experiment}

SVD Collaboration \cite{9} presented the preliminary results of a
search for the $\Theta^+$ in proton-nuclei(C,Si,Pb) interactions  at
70 GeV($\sqrt{s}$=11.5 GeV) IHEP accelerator (Protvino) with the
SVD-2 setup in the reaction chain: $pN\rightarrow \Theta^+X, \ \
\Theta^+ \rightarrow pK^0_s, \ \ K^0_s \rightarrow \pi^+\pi^-$ (SVD-2
experimental setup described in the paper). The data analysis was
done in the inclusive reactions with the limited multiplicity in the
proton beam fragmentation region. The resulting $pK^0_s$ effective
mass distribution before the cuts is shown in Fig. 10. The authors
point out that no obvious structure in this spectrum is seen, however
there is a small enhancement in the 1530 MeV mass region. After
applying a special cuts (details of cuts procedure can be found in
original paper) a narrow peak in the $pK^0_s$ effective mass
distribution was seen at the mass $M=1526\pm 3$ MeV with a
$\sigma=10\pm 3$ MeV. The statistical significance of this peak was
estimated to be of $5.6~\sigma$.

As it should be, we have also plotted in Fig. 10 the spectral lines
corresponding to KK excitations in the $pK^0$ system taken from Table
1. Apart from strong correlation of the spectral lines with the peaks
on the histogram in Fig. 10, here is found out too that the above
mentioned ``small enhancement in the 1530 MeV mass region" just
corresponds to the spectral line $M_6^{pK^0}$(1527) in KK tower for
the $pK^0$ system (see Table 1).

\subsection{ZEUS Collaboration experiment}

Unfortunately, we didn't have in hand the published paper of ZEUS
Collaboration. However, we can arrive at Ref. \cite{8} by $<$www$>$
where the preliminary results of an evidence for exotic baryon
decaying to $K_S^0$-(anti)proton were presented. A signal at
1527$\pm$2(stat) MeV with a Gaussian width of 10 $\pm$ 2 MeV and
statistical significance 4-5 $\sigma$ (from Gaussian fit) has been
observed. The measured mass is in excellent agreement with the
predicted state $M_6^{pK^0}$(1527) from KK tower for the $pK^0$
system (see Table 1). What is remarkable here that ZEUS Collaboration
observed this state for both $K^0$-proton and $K^0$-antiproton
channels as we predicted too.

Recently ZEUS Collaboration papers arrived at LANL e-print archive
\cite{13,14} where we found that a resonance search has been made in
the $K_S^0p$ and $K_S^0\bar p$ invariant mass spectrum measured with
the ZEUS detector at HERA using an integrated luminosity of 121
pb$^{-1}$, taken between 1996 and 2000. The $K_S^0p(\bar p)$
invariant mass spectrum has been studied in the central rapidity
region of inclusive deep inelastic scattering at an $ep$
center-of-mass energy of 300--318 GeV for a large range in the
exchanged-photon virtuality, $Q^2$, above 1 GeV$^2$. The results
presented there support the existence of exotic baryon decaying to
$K_S^0$-(anti)proton, and these results constitute first evidence for
the production of such a state in a kinematic region where hadron
production is dominated by fragmentation. The peak position,
determined from a fit to the mass distribution in the kinematic
region $Q^2\ge 20$ GeV$^2$, is $1521.5\pm 1.5({\rm
stat.})^{+2.8}_{-1.7} ({\rm syst.})$ MeV, and the measured Gaussian
width of $\sigma=6.1\pm 1.6({\rm stat.})^{+2.0}_{-1.4}({\rm syst.})$
MeV is above, but rather close to the experimental resolution of
$2.0\pm 0.5$ MeV. The signal is visible at high $Q^2$ and, for $Q^2 >
20$ GeV$^2$, contains $221 \pm 48$ events. The statistical
significance, estimated from the number of events assigned to the
signal by the fit, varied between $3.9\sigma$ and $4.6\sigma$
depending upon the treatment of the background. The mass of the
$\Theta^+$ baryon resonance reported now lies somewhat below the
average mass of the previous preliminary results \cite{8}. This is
because some additional cuts (energy and K* cuts) were removed in the
final analysis compared to the previous preliminary results, and only
dE/dx cuts were saved to get a good purity for the proton sample. The
ZEUS final results \cite{13,14} agree still with the ZEUS preliminary
results \cite{8} (without systematics) though.\footnote{I thank
S.~Chekanov for clarification of this point.}

The result of the fit using two Gaussians is shown in Fig.~11. It was
found that the second Gaussian significantly improved the fit in the
low mass region. It has a mass of $1465.1\pm 2.9 (\mathrm{stat.})$
MeV and a width of $15.5\pm 3.4 \mathrm{(stat.)}$ MeV and may
correspond to the $\Sigma (1480)$. However, as pointed out, the
parameters and significance of any state in this region are difficult
to estimate due to the steeply falling background close to threshold.
The probability of the 1522 MeV signal anywhere in the range
1500--1560 MeV arising from statistical fluctuation of the background
was below $6 \times 10^{-5}$. For a more realistic case, when the
starting background distribution included the 1465 MeV Gaussian the
probability was found to be about a factor of ten lower.

In order to determine the natural width of the state, the
Breit-Wigner function convoluted with the Gaussian was used in the
fitting procedure to describe the peak near 1522 MeV. If the width of
the Gaussian was fixed to the experimental resolution, the estimated
intrinsic width of the signal turned out $\Gamma=8\pm
4(\mathrm{stat.})$ MeV. The systematical error on this width was
expected to be smaller than the statistical error, but due to low
statistics, complicated background and the narrowness of the peak
leading to unstable fits, full systematical uncertainty was difficult
to estimate \cite{14}.

The invariant mass spectrum was also investigated for the $K_S^0p$
and $K_S^0\bar p$ samples separately. This result is shown as an
inset in Fig.~11 for $Q^2 > 20$ GeV$^2$. The results for two decay
channels are compatible, though the number of $K_S^0\bar p$
candidates is systematically lower.  The mass distributions were
fitted using the same function as the combined sample and gave
statistically consistent results for the peak position and width. The
number of events in the $K_S^0\bar p$ channel is $96 \pm 34$. If one
considers the signal in the $K_S^0p$ channel as the $\Theta^+$ then
the signal in the $K_S^0\bar p$ channel might be identified with the
$\Theta^-$. This is quite a new, remarkable observation made by ZEUS
Collaboration.

Besides, the $K^\pm p (K^\pm\bar{p})$ invariant mass spectra were
investigated for a wide range of minimum $Q^2$ values as well,
identifying proton and charged kaon candidates using $dE/dx$ in a
kinematic region similar to that used in the $K_S^0p(\bar p)$
analysis. For $Q^2>1$ GeV$^2$, no peak was observed near 1522 MeV in
the $K^+p$ spectrum, while a clean $10\sigma$ signal was observed in
the $K^-p (K^+\bar{p})$ channel at $1518.5\pm 0.6(\mathrm{stat.})$
MeV, corresponding to the PDG $\Lambda(1520)D_{03}$ state; see Fig.~5
in \cite{14}.

The $K^+p$ and $K^-p$ mass spectra were especially investigated at
$Q^2>20$ GeV$^2$, where the 1522 MeV peak is clearly pronounced for
the $K_S^0p(\bar p)$ channel. Again, no sign of a peak was found in
the $K^+p$ decays; see Fig.~6 in \cite{14}.

However, we would like to emphasize, looking at the Figures 5 and 6
in \cite{14}, one cannot definitely conclude that no structures are
seen in the $K^+p$ mass spectra. Certainly, more statistics and
higher mass resolution are needed to make a physically adequate
statement.

We have plotted in Fig.~11 the spectral lines corresponding to KK
excitations in the $pK^0$ system taken from Table 1. As seen in
Fig.~11, the spectral lines of the $M_3^{pK^0}(1460)$,
$M_4^{pK^0}(1477)$ and $M_5^{pK^0}(1500)$ from Kaluza-Klein tower for
kaon-nucleon system may correspond to the observed structures in
invariant mass spectra for the $K_S^0p(\bar p)$ channel.
Unfortunately, the statistics is small and mass resolution is not so
high to make a more reliable statement concerning the structures seen
from the left side of the 1522 MeV peak. We also found a strong
correlation of the calculated spectral lines with the enhancements on
the right side from the 1522 MeV peak in the invariant mass spectrum.
In this respect, it would be extremely important to improve the
statistics and mass resolution of the experiment, below and above of
the 1522 MeV peak, to confirm the lower and higher new exotic states
predicted.

\subsection{DUBNA propane bubble chamber experiment}

Quite an interesting results came just recently from DUBNA experiment
\cite{15} where the 2m propane bubble chamber experimental data have
been analyzed to search for the exotic baryon decaying to
$K_S^0$-proton in the reaction p+$C_3H_8$ at 10 GeV. The observation
of three peaks in the $pK^0_s$ invariant mass spectrum has been
reported with the masses $M_{K_S^0 p}$ = (1545.1$\pm$12.0,
1612.5$\pm$10.0, 1821.0$\pm$11.0) MeV and the measured widths of
$\Gamma_{K_S^0 p}$ = (16.3$\pm$3.6, 16.1$\pm$4.1, 28.0$\pm$9.4) MeV.
The statistical significance of these peaks were estimated to be of
(5.5$\pm$0,5)$\sigma$, (4.6$\pm$0.5)$\sigma$ and
(6.0$\pm$0.6)$\sigma$ respectively.

The effective mass distribution for 2300 $pK^0_S$ combination is
shown in Fig.~12. The background events were analyzed using the same
experimental condition. The background distribution was fitted by
six-order polynomial. The solid curve in Fig.~12 represents the sum
of the background and 4 Breit-Wigner resonances. As seen in Fig.~12,
there are significant enhancements in 1539, 1610 and 1810 MeV mass
regions. There is also a small peak in 1980 MeV mass region.

The total (5554 combination) $pK^0_S$ effective mass distribution is
shown in Fig.~13. The total experimental background was obtained with
the same experimental condition.  The significant enhancements are
clear seen in Fig.~13 in 1545, 1616 and 1811 MeV mass regions. A
small peak in 2.0 GeV mass region is pointed out too.

The calculated spectral lines corresponding to KK excitations in the
$pK^0$ system taken from Table 1 have been plotted in Figs. 12 and 13
as well. Here, as in previous cases, we also found a remarkable
coincidence of the calculated spectral lines with the enhancements in
the invariant mass spectra.

\section{Summary and Discussion}

We have presented an analysis of the recent results from several
experimental groups reported observation of a new $\Theta$ baryon.
Our analysis was motivated by the wishes to understand a real origin
of the $\Theta$ baryon and to find its place in the unified picture
for hadron spectra developed early \cite{12}. A thorough analysis of
all experiments mentioned in Introduction taken together shows that,
in fact, two different $\Theta$ baryons have been discovered: one
$\Theta^\pm$ baryon in the $p^\pm K^0$ system and other $\Theta^+$
baryon in the $nK^+$ system. It's our first claim.

We have established that both $\Theta$ baryons are excellently
incorporated in the unified picture for hadron spectra: the
$\Theta(pK_S^0)$ baryon lives on the six storey in Kaluza-Klein tower
for $pK^0$ system but the $\Theta(nK^+)$ baryon lives on the seven
storey in Kaluza-Klein tower for $nK^+$ system; see Table 1.

As is seen from Table 1, almost all experimentally filled storeys in
Kaluza-Klein tower for kaon-nucleon system contain pairs of $\Sigma$
and $\Lambda$ particles with negative strangeness. One might think
that $\Theta(pK_S^0)$ baryon represents an isovector with positive
strangeness partner of isoscalar $\Lambda$(1520) particle with
negative strangeness, and $\Theta(nK^+$ baryon is isoscalar with
positive strangeness partner of isovector $\Sigma$(1560) particle
with negative strangeness. In this respect, it would be quite an
important to set up purposeful experiments to search for the states
$\Sigma(1520)\in M_6^{NK}(1524-1528)\bigcup
M_8^{\Lambda\pi}(1522-1524)$ and $\Lambda (1560)\in
M_7^{NK}(1555-1560)$. It would also be very interesting an attempt to
seek a rare decay mode $\Lambda(1520)\rightarrow \Lambda\pi$ with an
isospin violation (see 8-storey in Table 2). From strong correlation
between spectral lines corresponding to Kaluza-Klein tower for
kaon-nucleon system and the peaks on the histograms established
above, one can conclude that there exist positive srangeness $\Theta$
partners for all $\Sigma$ and $\Lambda$ particles with negative
strangeness as we predicted -- this is our second claim. The most
impressive manifestation and confirmation of that claim we found in
neutrino interactions data set \cite{5}. Our predictions should be
tested in the future experimental studies.

The properties of the observed $\Theta$ states, such as spin,
isospin, and parity are not established in the above mentioned
experiments. So, it is important to measure the quantum numbers of
the $\Theta$ particles in the future experiments. In our opinion, a
classification of baryons by soliton configurations in chiral field
theories is more preferable compared to the standard quark model
classification. In particular, the SU(3) version of the chiral
soliton model \cite{10} predicts the $\Theta^+$ state of having spin
1/2, isospin 0, and positive parity. Attractive in the soliton
classification scheme that the $\Theta$ baryon is considered on an
equal ground with other baryons. This is ideologically close to our
scheme of systematics. However, our description significantly differs
in many important things from the chiral soliton description.

Our conservative estimate for the widths of KK excitations looks like
$\Gamma_n\sim 0.4\cdot n$ Mev, where $n$ is the number of KK
excitation. This gives $\Gamma_6(\Theta^\pm)\sim 2.4$ Mev and
$\Gamma_7(\Theta^+)\sim 2.8$ Mev. Thus, a more careful experimental
studies with a higher statistics and mass resolution are extremely
important.

Here, we should like to bring up some comments on the problems under
discussion. If one takes the structures below and above of the
$\Theta$ peak in the $K_S^0p$ invariant mass spectra observed in
\cite{13,14,15} seriously, then we come to the conclusion that the
experimentally observed picture is not at all what is predicted by
the chiral soliton model. As an additional argument in favor of that
claim we would like to refer to the recent, interesting article
\cite{16} where a careful analysis of $K^+d$ total cross section data
was undertaken to explore possible manifestations of the resonance in
the S=+1 hadronic system with mass around 1.55 GeV. It was found that
a structure corresponding to the resonance is visible in the data.
The width consistent with the observed deviation from background was
found to be 0.9$\pm$0.2 MeV and the mass is 1.559$\pm$0.003 GeV for
the spin parity ${\frac{1}{2}}^+$ and 1.547$\pm$0.002 GeV for
${\frac{1}{2}}^-$. Perhaps this is the first experimental verdict on
a crucial problem concerning the quantum numbers of the $\Theta$
baryon. The obtained mass 1.559$\pm$0.003 GeV for the spin parity
${\frac{1}{2}}^+$ is in excellent agreement with the
$M_7^{nK^\pm}$(1556\,MeV)-storey in Kaluza-Klein tower for
kaon-nucleon system (see Table 1) and far enough from accurate
prediction 1530 MeV  in \cite{10}. The mass 1.547$\pm$0.002 GeV is
closer to the prediction 1530 MeV but the negative parity
(${\frac{1}{2}}^-$) in that case contradicts to the predicted
\cite{10} positive parity (${\frac{1}{2}}^+$). We would also like to
point out, once again, a noticeable spread among the masses of the
$\Theta$ states observed in different experiments. However, it should
be noted that the CLAS Collaboration \cite{6} reported a very small
statistical error on the mass of the $\Theta^+$ which is
1555$\pm$1(stat.) MeV. The large error in the quoted mass \cite{6}
was due to possible systematic errors. The mass value
1555$\pm$1(stat.) MeV is in excellent agreement with the
$M_7^{nK^\pm}$(1556\,MeV)-storey in the Kaluza-Klein tower for
kaon-nucleon system as well, and this might be considered as an
additional argument in favor of the spin parity ${\frac{1}{2}}^+$.
Certainly, this is a remarkable fact that it may be possible to infer
the parity of the state from a comparison of its mass values
\cite{16}. In this respect a more accurate experimental measurement
and precise determination of the masses of the $\Theta$ states is a
question of vital importance as well. No doubt, further measurements
to obtain a more accurate kaon-nucleon scattering data would be very
useful to better establish all possible states in kaon-nucleon system
with positive and negative strangeness simultaneously.

We have already claimed that the year 2003 will enter the history of
particle physics as a year of fundamental discoveries \cite{12}. This
concerned a series of new narrow mesons states discovered at the
masses which are surprisingly far from the predictions of
conventional quark potential models. Exciting experimental
measurements in baryons spectroscopy in the same year and next with
discovery of the $\Theta$ baryon states provided an additional
excellent confirmation of that claim. All of these discoveries in
meson and baryon spectroscopy clearly open a new page in hadron
physics which is certainly a starting page of a new era in particle
physics.  One might think that this is an era of the extra
dimensions.

\section*{Acknowledgements}

I would like to thank A.V.~Razumov for friendly table conversations
and useful professional advices.

\newpage
\vspace*{2cm}
\begin{center}
Table 1: Kaluza-Klein tower of KK excitations for kaon-nucleon system
and experimental data. The $\Theta^+$ mass values have been taken
from Refs. \cite{6,7}.

\vspace{5mm}
\begin{tabular}{|c|c|c|c|c|l|}\hline
 n & $ M_n^{nK^{\pm}}$MeV & $ M_n^{pK^{0}}$MeV & $ M_n^{nK^{0}}$MeV
 & $ M_n^{pK^{\pm}}$MeV & $ M_{exp}^{NK}$\,MeV \\
 \hline
1  & 1435.90 & 1438.59 & 1439.88 & 1434.61 & \\
2  & 1443.82 & 1446.47 & 1447.76 & 1442.53 & \\
3  & 1456.89 & 1459.48 & 1460.77 & 1455.61 & \\
4  & 1474.92 & 1477.43 & 1478.71 & 1473.65 & $\Sigma$(1480)\\
5  & 1497.66 & 1500.08 & 1501.35 & 1496.40 & \\
6  & 1524.84 & 1527.16 & 1528.41 & 1523.59 & $\Lambda$(1520)\,\Tetat \\
7  & 1556.15 & 1558.36 & 1559.60 & 1554.92 & $\Sigma$(1560)\,\Tetas \\
8  & 1591.30 & 1593.39 & 1594.61 & 1590.07 & $\Sigma$(1580)\,$\Lambda$(1600)\\
9  & 1629.96 & 1631.95 & 1633.16 & 1628.76 & $\Sigma$(1620)\,$\Lambda$(1600)\\
10 & 1671.87 & 1673.76 & 1674.94 & 1670.69 & $\Sigma$(1670)\,$\Lambda$(1670)\\
11 & 1716.75 & 1718.53 & 1719.69 & 1715.59 & $\Sigma$(1690)\,$\Lambda$(1690)\\
12 & 1764.35 & 1766.02 & 1767.17 & 1763.20 & $\Sigma$(1770)\,$\Lambda$(1800)\\
13 & 1814.42 & 1816.01 & 1817.13 & 1813.30 & $\Sigma$(1840)\,$\Lambda$(1830)\\
14 & 1866.77 & 1868.26 & 1869.36 & 1865.67 & $\Sigma$(1880)\,$\Lambda$(1890)\\
15 & 1921.19 & 1922.60 & 1923.68 & 1920.11 & $\Sigma$(1915)\\
16 & 1977.51 & 1978.85 & 1979.90 & 1976.46 & $\Sigma$(1940)\,$\Lambda$(2000)\\
17 & 2035.57 & 2036.84 & 2037.87 & 2034.54 & $\Sigma$(2030)\,$\Lambda$(2020)\\
18 & 2095.23 & 2096.43 & 2097.44 & 2094.22 & $\Sigma$(2080)\,$\Lambda$(2100)\\
19 & 2156.35 & 2157.49 & 2158.48 & 2155.36 & $\Sigma$(2100)\,$\Lambda$(2110)\\
20 & 2218.82 & 2219.90 & 2220.87 & 2217.85 & $\Sigma$(2250)\\
21 & 2282.52 & 2283.55 & 2284.50 & 2281.57 & \\
22 & 2347.37 & 2348.35 & 2349.28 & 2346.44 & $\Lambda$(2350)\\
23 & 2413.27 & 2414.20 & 2415.11 & 2412.36 & \\
24 & 2480.14 & 2481.03 & 2481.92 & 2479.25 & \\
25 & 2547.91 & 2548.76 & 2549.63 & 2547.04 & $\Lambda$(2585)\\
26 & 2616.50 & 2617.32 & 2618.17 & 2615.66 & $\Sigma$(2620)\\
27 & 2685.88 & 2686.66 & 2687.49 & 2685.04 & \\
28 & 2755.96 & 2756.72 & 2757.53 & 2755.15 & \\
29 & 2826.71 & 2827.44 & 2828.23 & 2825.92 & \\
30 & 2898.08 & 2898.78 & 2899.56 & 2897.30 & \\ \hline
\end{tabular}
\end{center}

\newpage

\begin{center}
Table 1: Kaluza-Klein tower of KK excitations for $\Lambda\pi$ system
and experimental data.

\vspace{5mm} {\large
\begin{tabular}{|c|c|c|l|}\hline
 n & $ M_n^{\Lambda\pi^0}$MeV & $ M_n^{\Lambda\pi^\pm}$MeV & $
 M_{exp}^{\Lambda\pi}$MeV   \\
 \hline
1  & 1257.66 & 1262.06 &  \\
2  & 1277.20 & 1281.13 &  \\
3  & 1306.19 & 1309.59 &  \\
4  & 1341.85 & 1344.77 &  \\
5  & 1382.26 & 1384.79 & $\Sigma$(1385) \\
6  & 1426.24 & 1428.46 &  \\
7  & 1473.06 & 1475.02 & $\Sigma$(1480) \\
8  & 1522.24 & 1524.00 & $\Sigma(\Lambda)$(1520)?? \\
9  & 1573.47 & 1575.06 & $\Sigma$(1560) \\
10 & 1626.52 & 1627.96 & $\Sigma$(1620) \\
11 & 1681.22 & 1682.55 & $\Sigma$(1670) \\
12 & 1737.44 & 1738.66 & $\Sigma$(1750) \\
13 & 1795.06 & 1796.20 & $\Sigma$(1840) \\
14 & 1853.99 & 1855.05 & $\Sigma$(1880) \\
15 & 1914.15 & 1915.14 & $\Sigma$(1915) \\
16 & 1975.46 & 1976.39 & $\Sigma$(1940) \\
17 & 2037.84 & 2038.72 & $\Sigma$(2030) \\
18 & 2101.24 & 2102.07 & $\Sigma$(2080) \\
19 & 2165.60 & 2166.39 & $\Sigma$(2100) \\
20 & 2230.87 & 2231.62 & $\Sigma$(2250) \\
21 & 2296.98 & 2297.69 &  \\
22 & 2363.89 & 2364.57 &  \\
23 & 2431.56 & 2432.21 &  \\
24 & 2499.94 & 2500.57 &  \\
25 & 2568.99 & 2569.59 &  \\
26 & 2638.68 & 2639.26 &  \\
27 & 2708.96 & 2709.52 &  \\
28 & 2779.81 & 2780.35 &  \\
29 & 2851.19 & 2851.71 &  \\
30 & 2923.07 & 2923.58 &  \\ \hline
\end{tabular}}
\end{center}

\newpage

\begin{figure} [htb]
\begin{center}
\includegraphics[width=\textwidth]{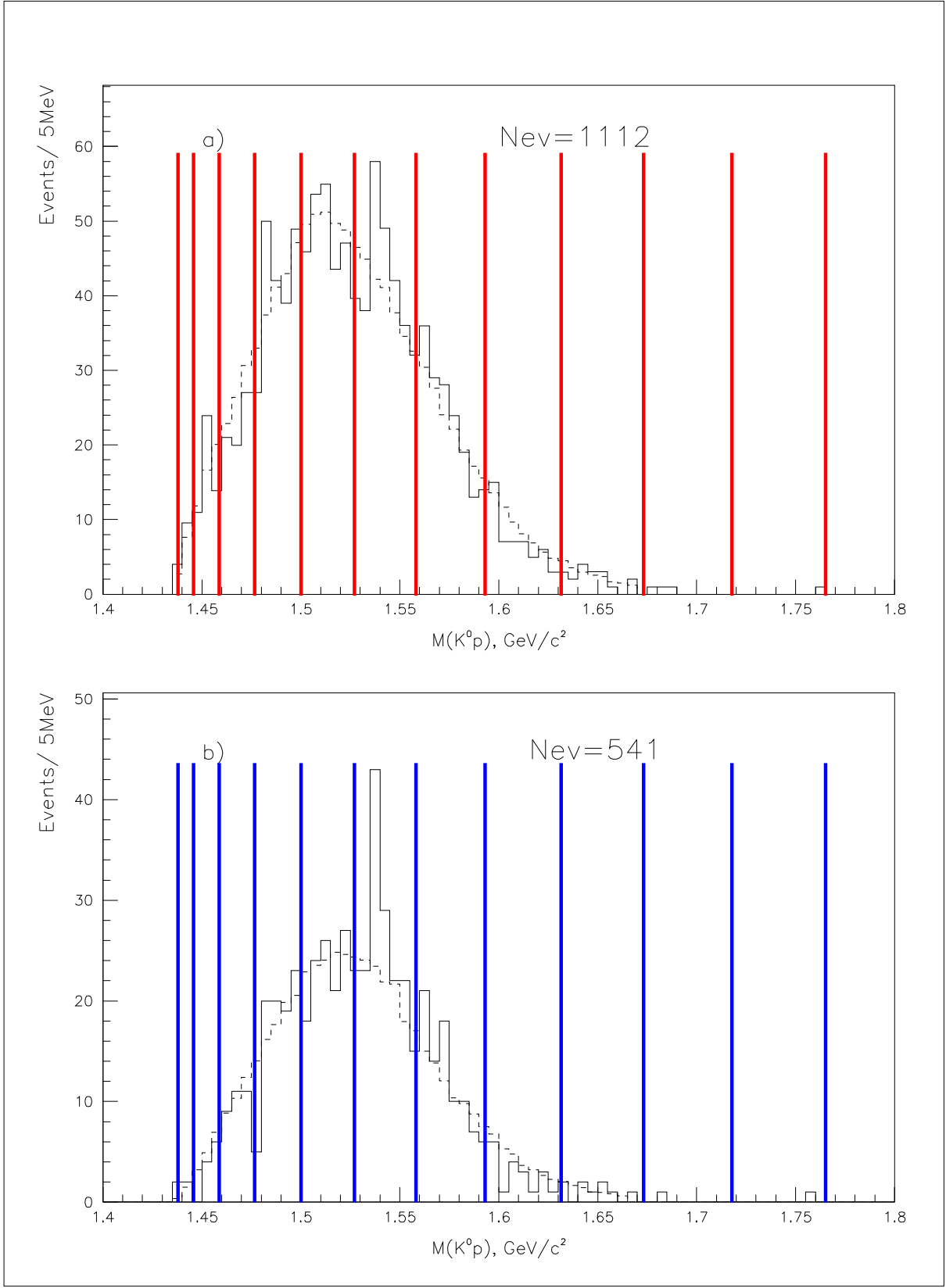}\label{fig1}
\end{center}

\caption{Effective mass of the $K^0 p$ system formed in the reaction
\charex \ \  ex\-tracted from Ref. \cite{2}: (a) for all measured
events, (b) for events that pass additional selections; see \cite{2}.
The vertical (spectral) lines correspond to KK tower for $K^0 p$
system; see Table 1.}
\end{figure}

\newpage

\begin{figure}[htb]
\begin{center}
\includegraphics[width=\textwidth]{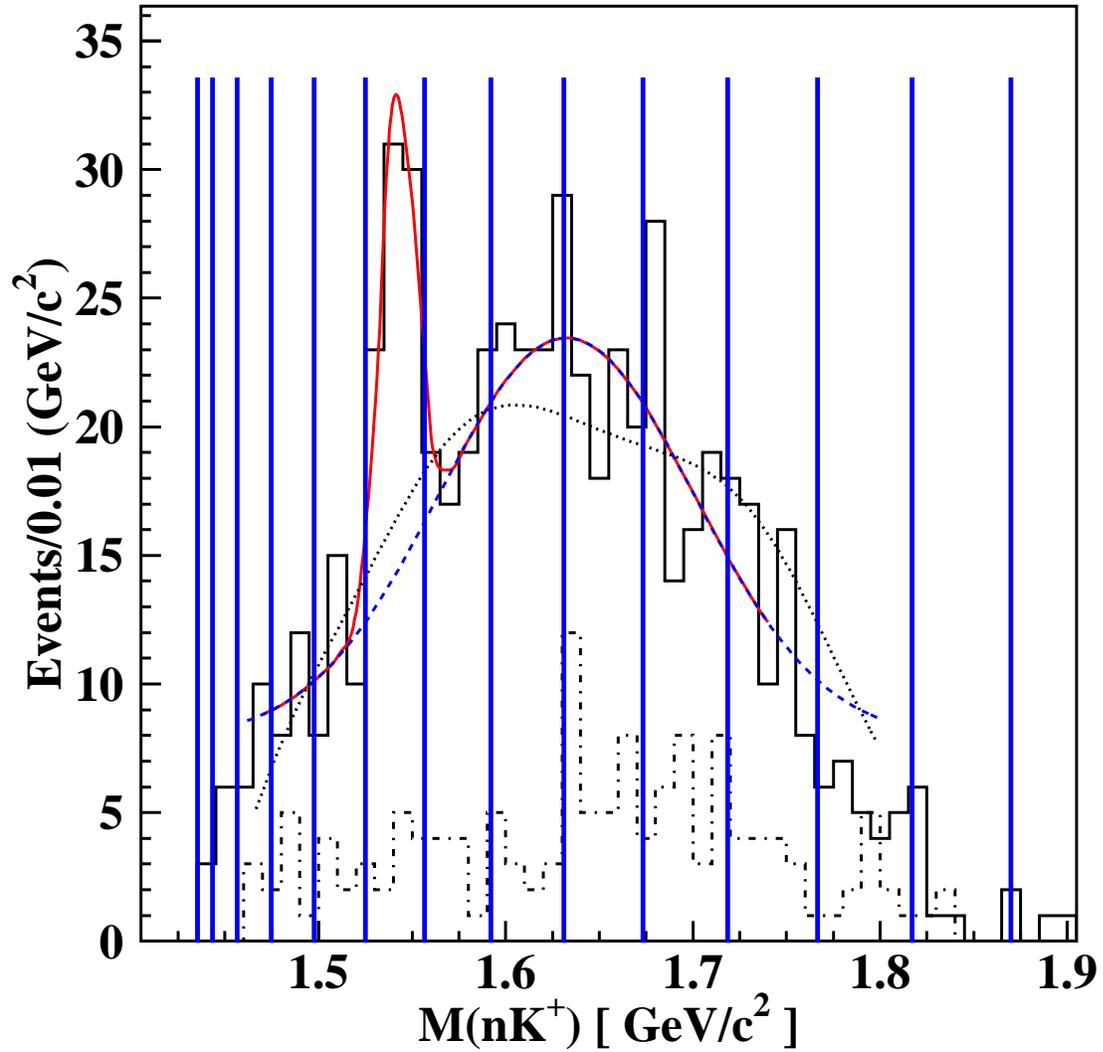}\label{fig2}
\caption{ Invariant mass of the $nK^+$ system extracted from Ref.
\cite{3}. The vertical (spectral) lines correspond to KK tower for
$nK^+$ system; see Table 1.}
\end{center}
\end{figure}

\newpage

\begin{figure}
\begin{center}
\includegraphics[width=\textwidth]{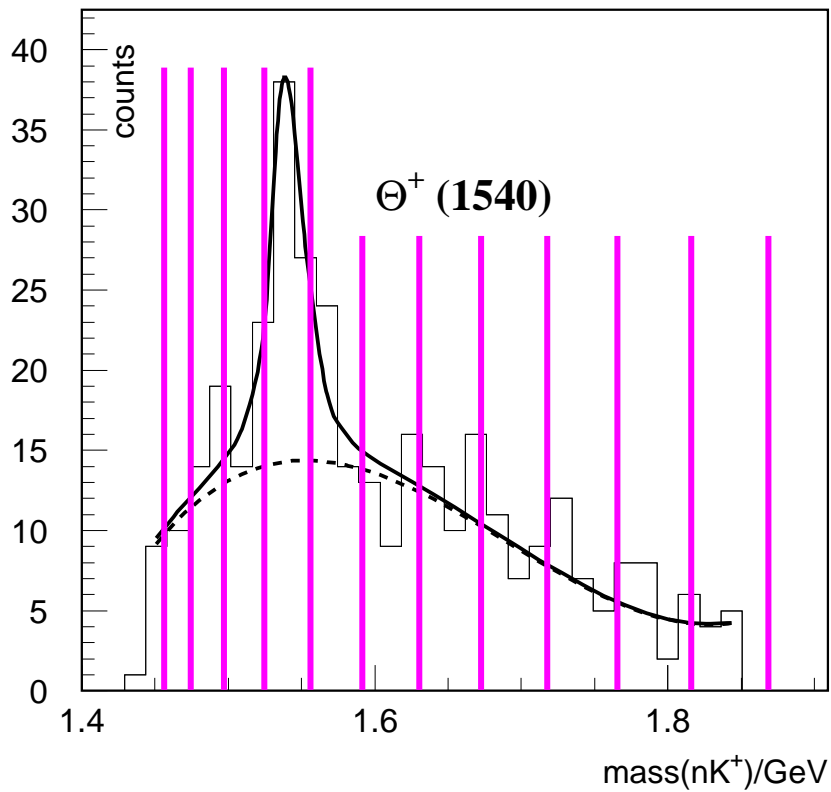}\label{fig3}
\vspace{-15mm}
\caption{The $nK^+$ mass distribution
after cuts in the $\pi^+\pi^-$ mass distribution and in the $K^0_S$
production angle extracted from Ref. \cite{4}. The solid line
represents a fit using a Breit--Wigner distribution (convoluted with
a resolution function) plus polynomial background; see \cite{4} for
details. The vertical (spectral) lines correspond to KK tower for
$nK^+$ system; see Table~1.}
\end{center}
\end{figure}

\newpage

\begin{figure}[htb]
\begin{center}
\includegraphics[width=\textwidth]{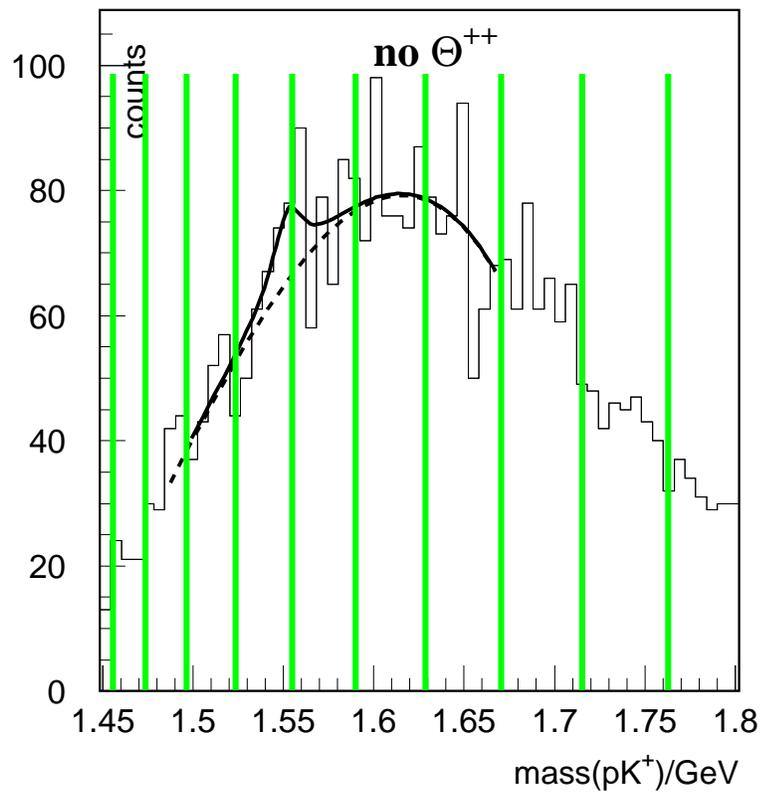} \label{fig4}
\vspace{-15mm}
\caption{The $pK^+$ invariant mass distribution in the
$pK^+K^-$ final state extracted from Ref. \cite{4}. The solid line
represents a fit using a Breit--Wigner distribution (convoluted with
a resolution function) plus polynomial background; see \cite{4} for
details. The vertical (spectral) lines correspond to KK tower for
$pK^+$ system; see Table~1.}
\end{center}
\end{figure}

\newpage

\begin{figure}[htb]
\begin{center}
\includegraphics[width=\textwidth]{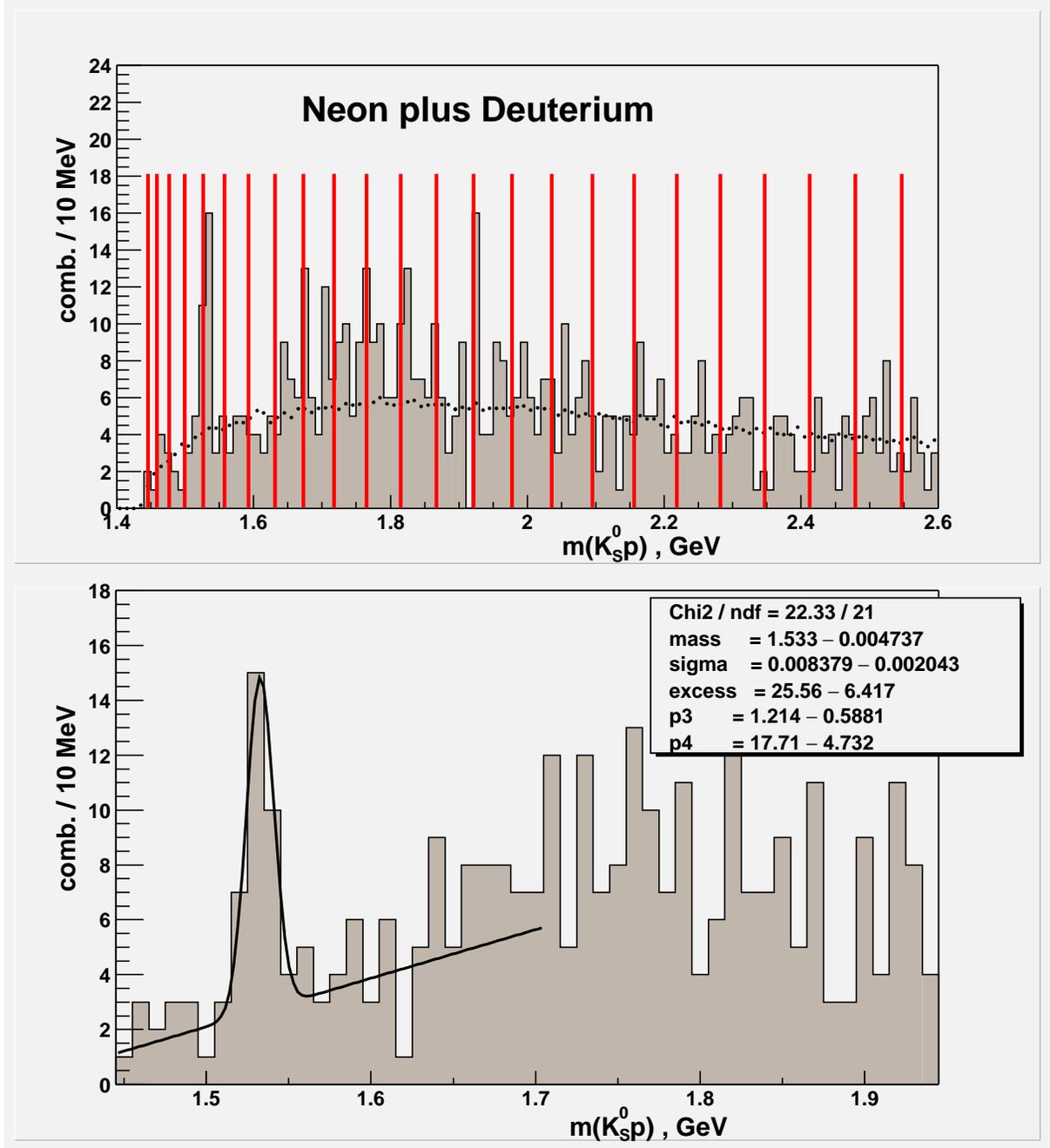}\label{fig5}
\caption {Extracted from Ref. \cite{5} the $K_S^0p$ invariant mass
distribution for the Neon and Deuterium data combined (top panel).
The dots depict the random-star background. The vertical (spectral)
lines correspond to KK tower for $pK^0$ system; see Table~1. A fit of
the same $K_S^0p$ distribution but plotted with shifted bins is shown
in the bottom panel; see \cite{5} for details.}
\end{center}
\end{figure}

\newpage

\begin{figure}[htb]
\begin{center}
\includegraphics[width=\textwidth]{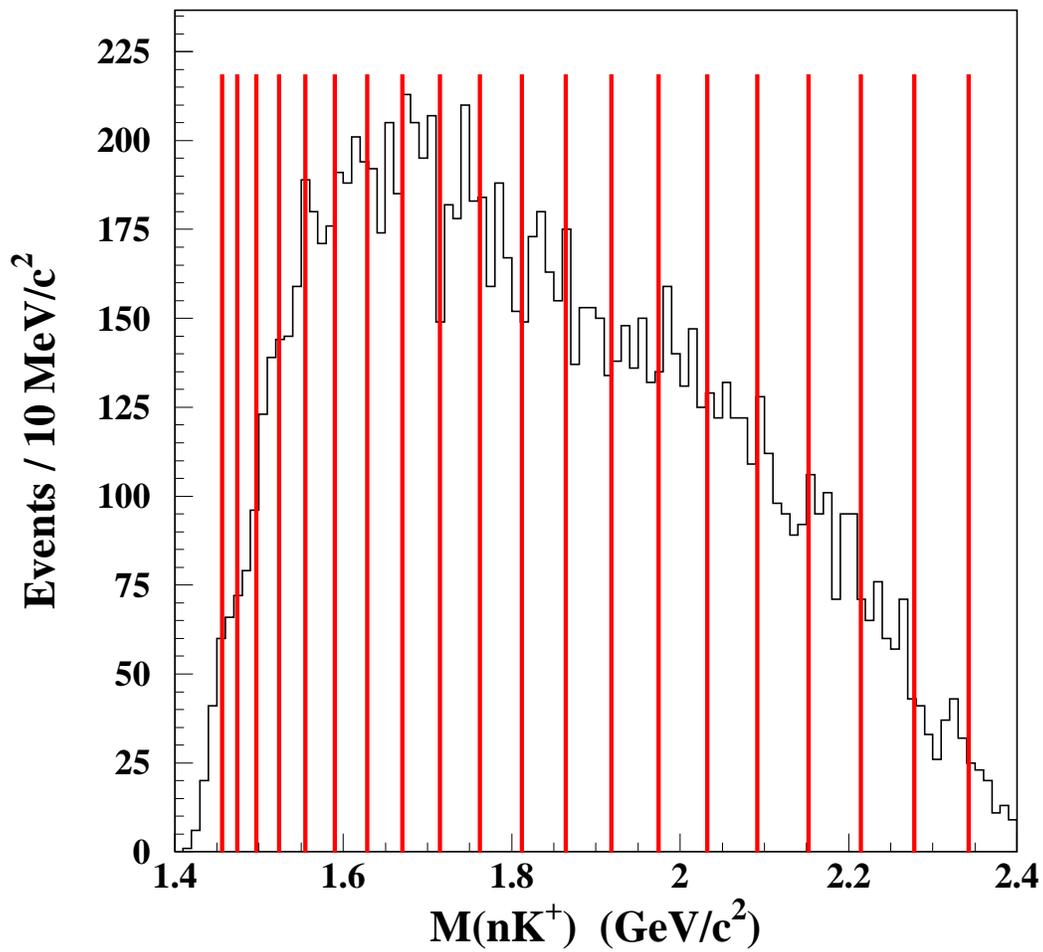}\label{fig6}
\vspace{-25mm} \caption{ The $n K^+$ invariant mass spectrum in the
reaction $\gamma p\rightarrow \pi^+K^-K^+(n)$ extracted from Ref.
\cite{6}.  The vertical (spectral) lines correspond to KK tower for
$nK^+$ system; see Table~1.}
\end{center}
\end{figure}

\newpage

\begin{figure}[htb]
\begin{center}
\includegraphics[width=\textwidth]{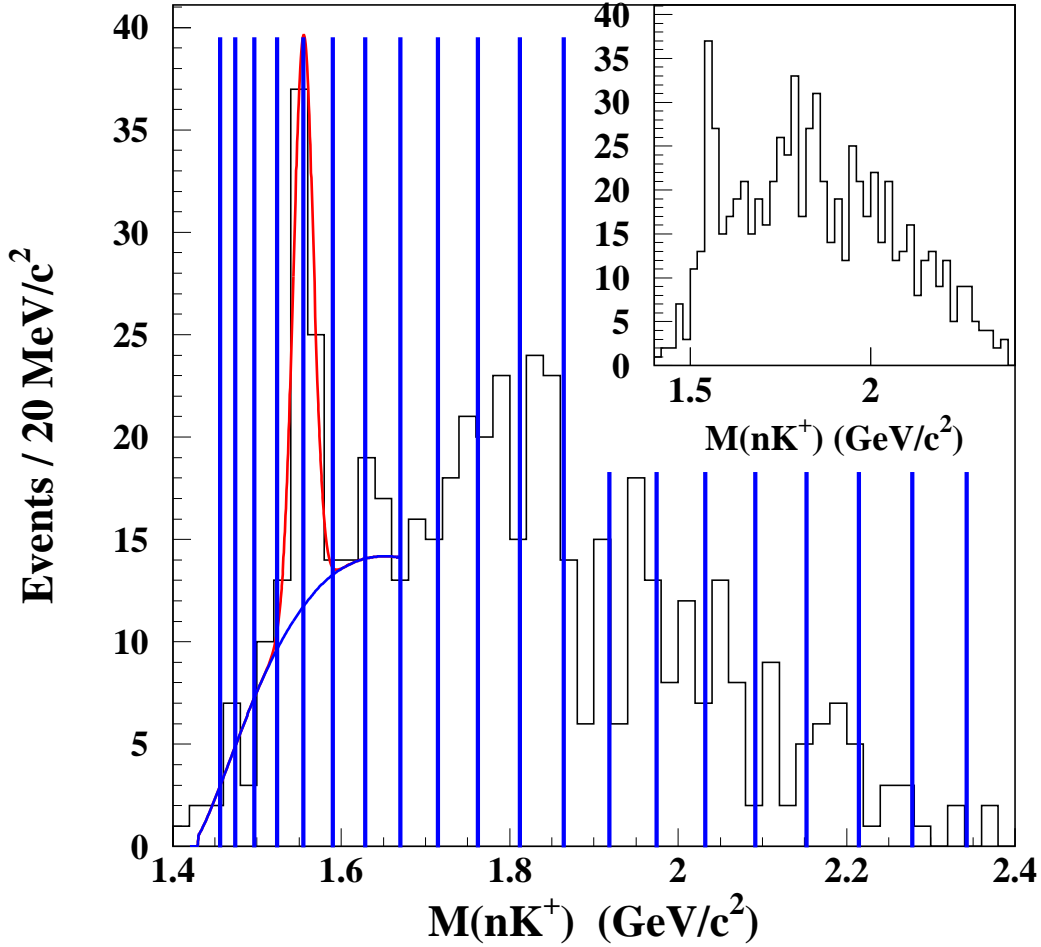}\label{fig7}
\vspace{-25mm} \caption{Extracted from Ref. \cite{6} the $nK^+$
invariant mass spectrum in the reaction $\gamma p\rightarrow
\pi^+K^-K^+(n)$ with the cut $\cos\theta^*_{\pi^+}>0.8$ and
$\cos\theta^*_{K^+}<0.6$. $\theta^*_{\pi^+}$ and $\theta^*_{K^+}$ are
the angles between the $\pi^+$ and $K^+$ mesons and photon beam in
the center-of-mass system. The vertical (spectral) lines correspond
to KK tower for $nK^+$ system; see Table~1. The inset shows the
$nK^+$ invariant mass spectrum with only the
$\cos\theta^*_{\pi^+}>0.8$ cut; see \cite{6} for details. }
\end{center}
\end{figure}

\newpage

\begin{figure} [htb]
\begin{center}
\includegraphics[width=.7\textwidth]{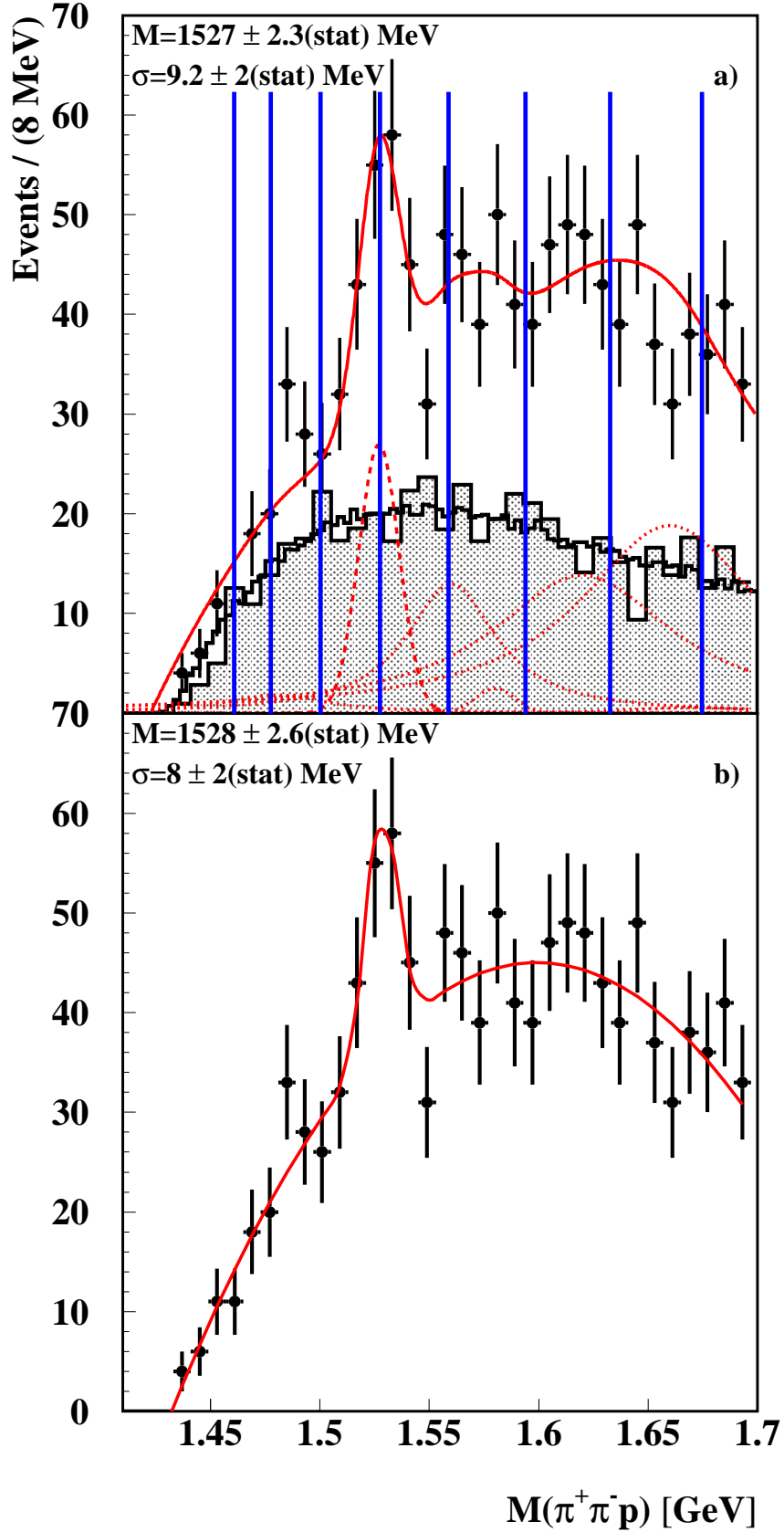}\label{fig8}
\end{center}
\vspace{-8mm} \caption{Distribution in invariant mass of the $p
\pi^+\pi^-$ system subject to various constraints described in Ref.
\cite{7}. In top panel the Phythia6 Monte Carlo simulation is
represented by the gray shaded histogram; see \cite{7} for details.
The vertical (spectral) lines in top panel correspond to KK tower for
$K^0p$ system; see Table~1.}
\end{figure}

\newpage

\begin{figure}[htb]
\begin{center}
\includegraphics[width=\textwidth]{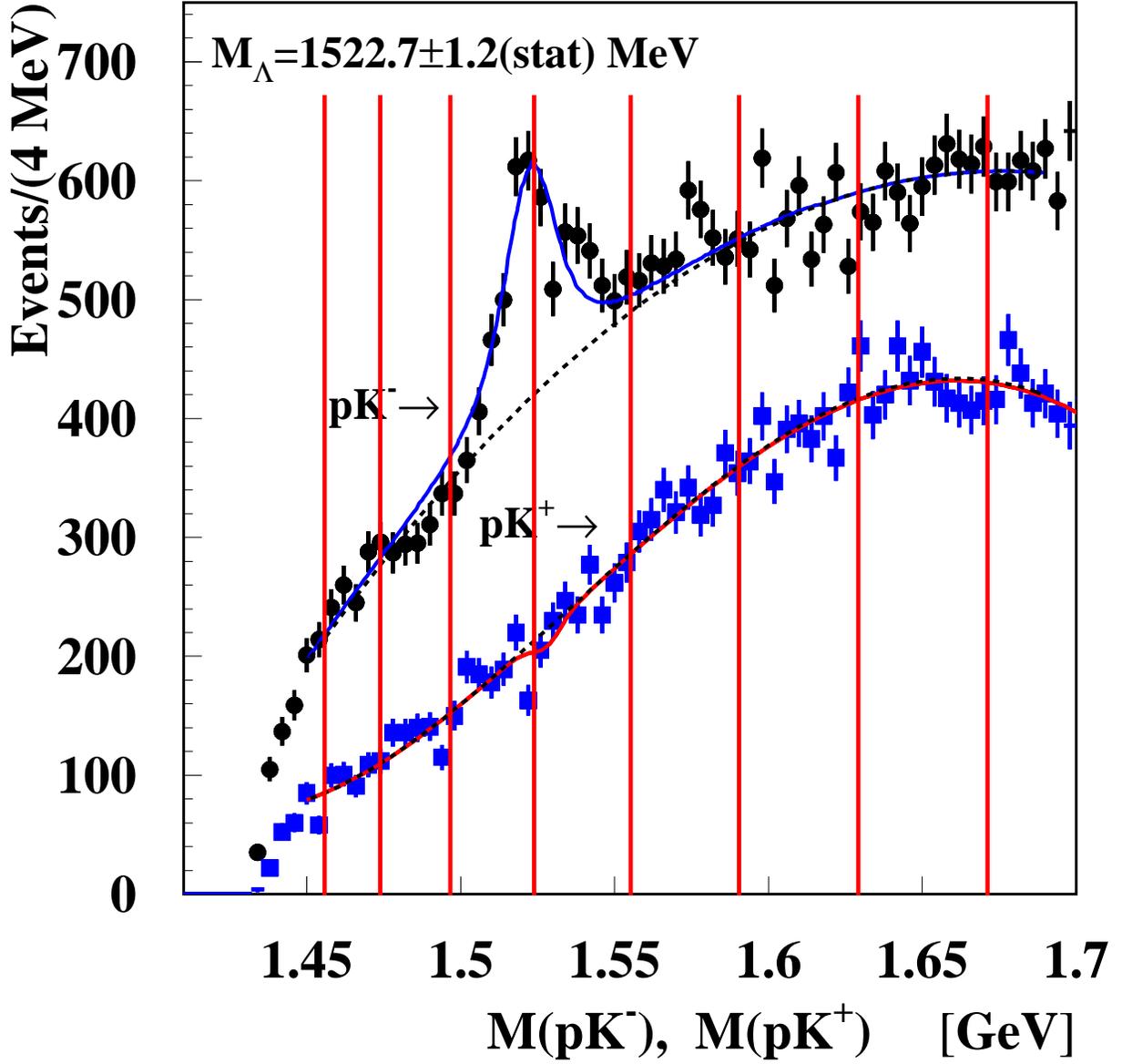}\label{fig9}
\end{center}
\vspace{-5mm} \caption{Spectra of invariant mass $M_{pK^-}$ (top) and
$M_{pK^+}$ (bottom) extracted from Ref. \cite{7}. The vertical
(spectral) lines
 correspond to KK tower for $K^\pm p$ system; see Table~1.}
\end{figure}

\newpage

\begin{figure}[htb]
\begin{center}
\includegraphics[width=\textwidth]{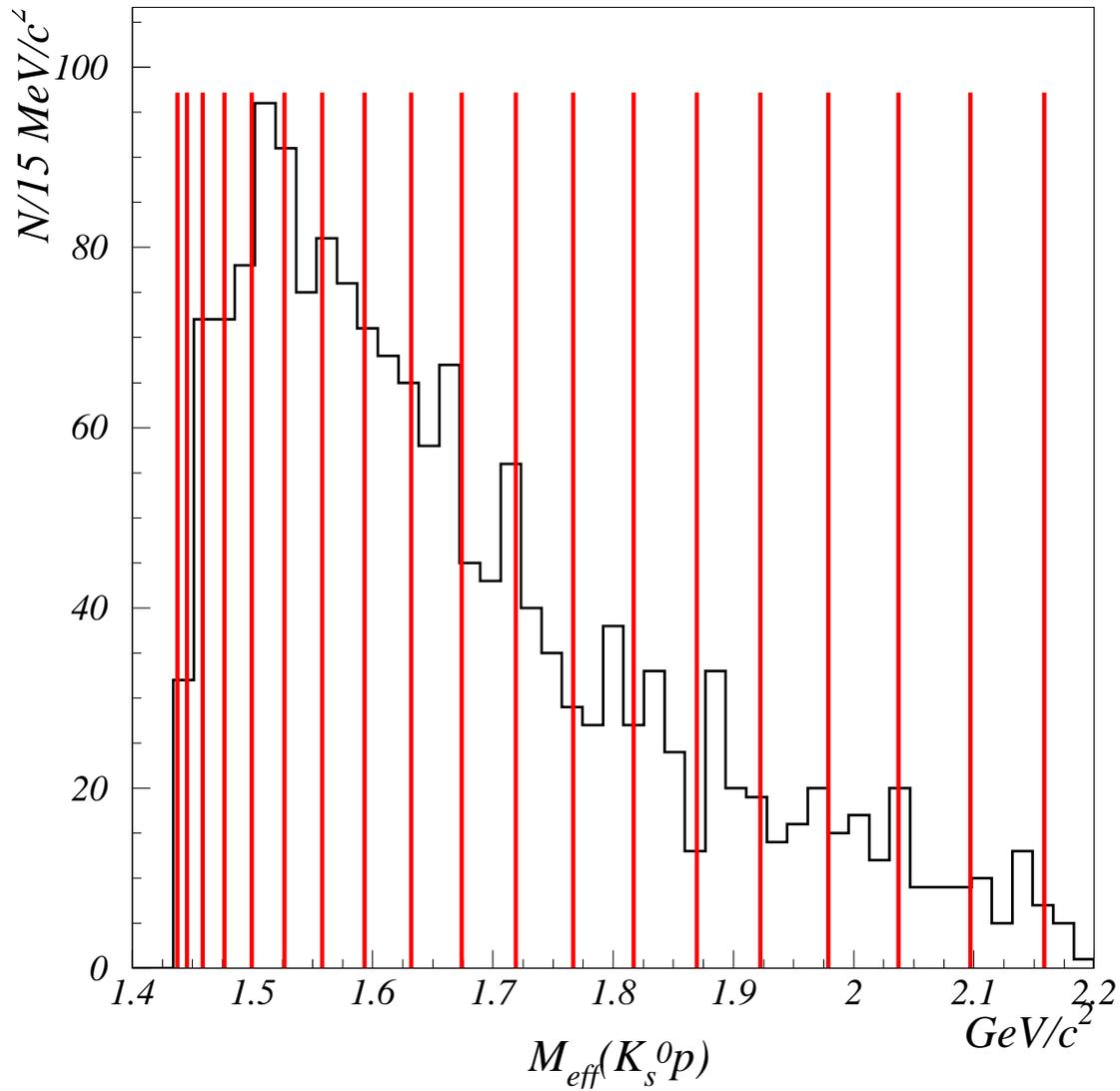}\label{fig10}
\end{center}
\vspace{-5mm} \caption{ The $(pK^0_S)$ invariant mass spectrum in the
reaction $pA\rightarrow pK^0_SA+X$ before the cuts extracted from
Ref. \cite{9}. The vertical (spectral) lines correspond to KK tower
for $K^0 p$ system; see Table~1.}
\end{figure}

\newpage

\begin{figure}[htb]
\begin{center}
\includegraphics[width=\textwidth]{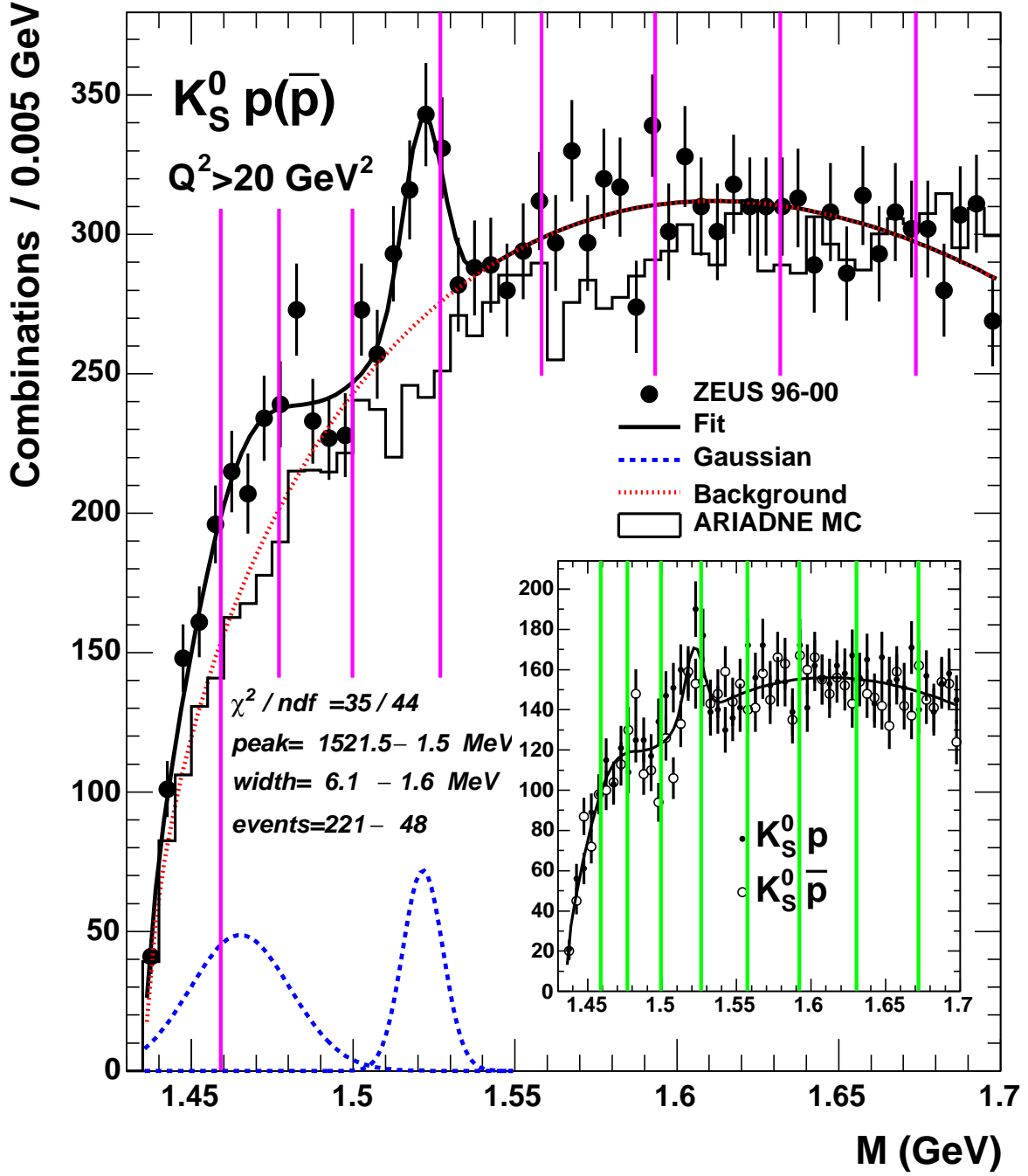}\label{fig11}
\end{center}
\vspace{-5mm} \caption{Invariant-mass spectrum for the $K_S^0p(\bar
p)$ channel for $Q^2 > 20$ GeV$^2$ extracted from Ref. \cite{13}. The
solid line is the result of a fit to the data using a three-parameter
background function plus two Gaussians (see \cite{13}). The dashed
lines show the Gaussian components and the dotted line the background
according to this fit. The histogram shows the prediction of the
ARIADNE MC simulation normalized to the data in the mass region above
1650 MeV. The inset shows the $K_S^0\bar p$ (open circles) and the
$K_S^0p$ (black dots) candidates separately, compared to the result
of the fit to the combined sample scaled by a factor of 0.5. The
vertical (spectral) lines correspond to KK tower for $K^0 p$ system;
see Table~1.}
\end{figure}

\newpage

\begin{figure}[htb]
\begin{center}
\includegraphics[width=\textwidth]{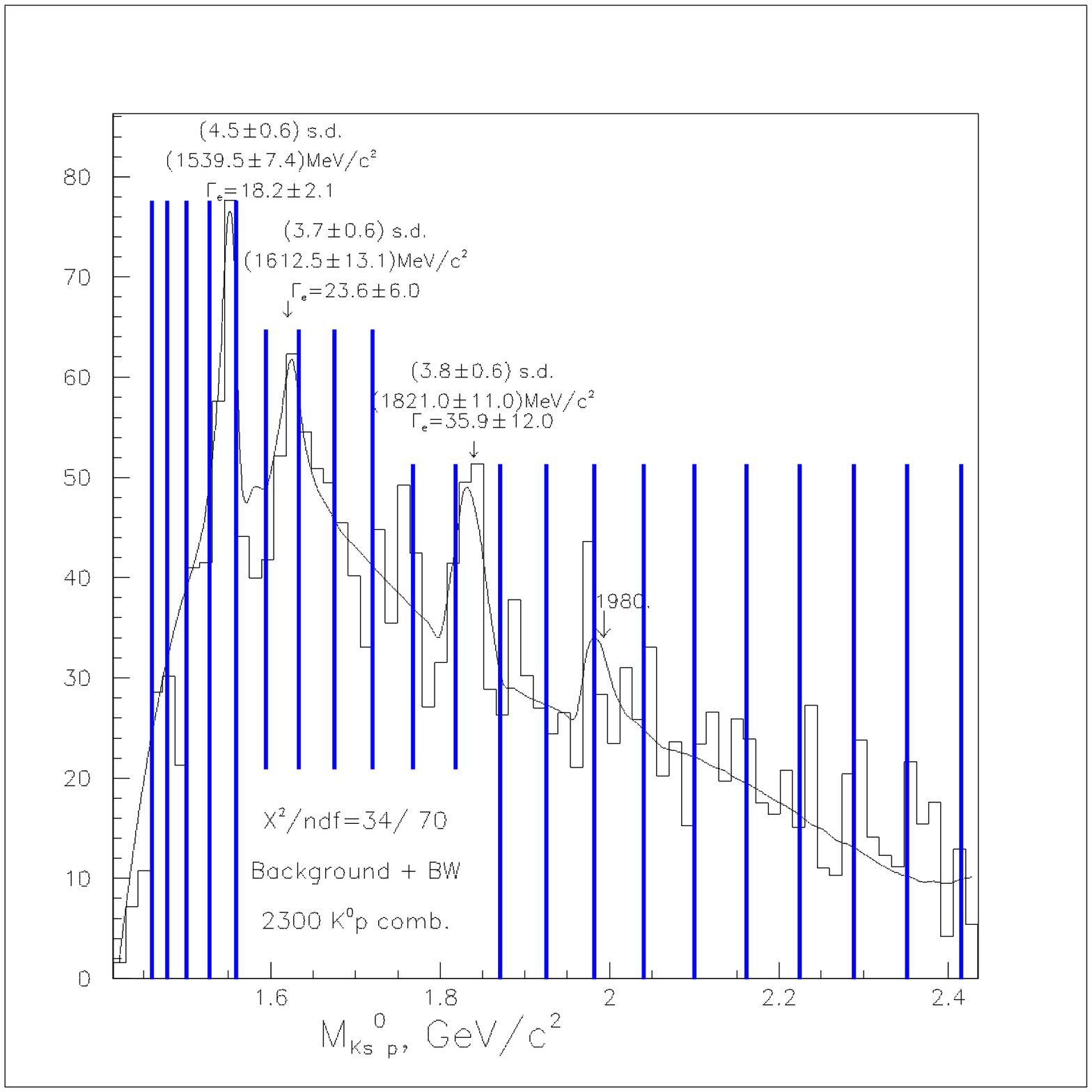}\label{fig12}
\end{center}
\vspace{-7cm} \caption{The $pK_S^0$ invariant-mass spectrum with the
momentum of $0.350\le p\le 0.900$GeV/c for identified protons in the
reaction p+$C_3H_8\to pK^0_s$ + X extracted from Ref. \cite{15}. The
solid curve represents the sum of the experimental background and 4
Breit-Wigner resonances. The experimental background was approximated
by six-order polynomial. The vertical (spectral) lines correspond to
KK tower for $K^0 p$ system; see Table~1.}
\end{figure}

\newpage

\begin{figure}[htb]
\begin{center}
\includegraphics[width=\textwidth]{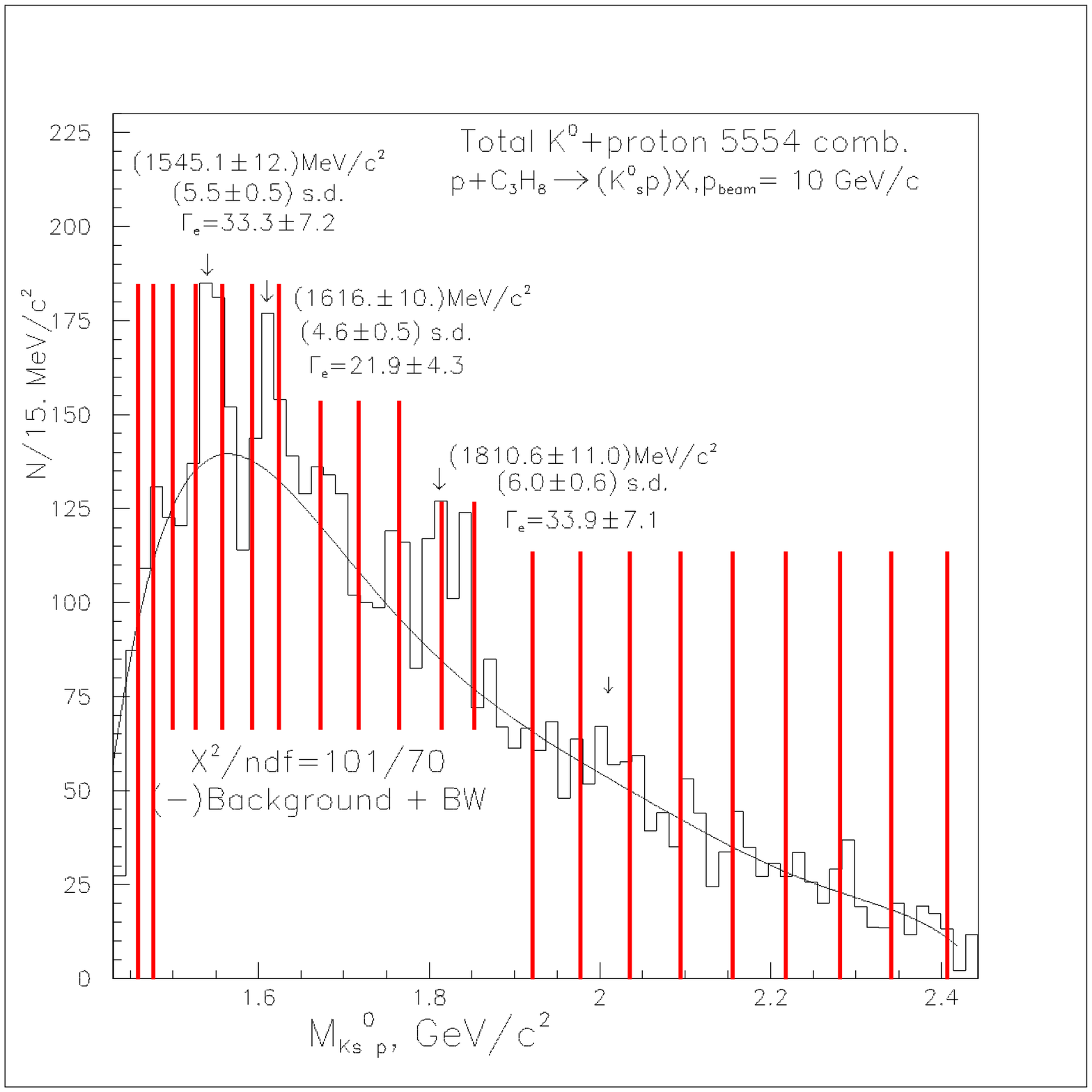}\label{fig13}
\end{center}
\vspace{-7cm} \caption{The $K^0_Sp$ effective mass distribution
combined for protons with the momenta $0.350\le p\le 0.9$ GeV/c$^2$
and $p\ge1.7 GeV/c^2$ extracted from Ref. \cite{15}. The curve is the
experimental background taken in the form of six-order polynomial.
The vertical (spectral) lines correspond to KK tower for $K^0 p$
system; see Table~1.}
\end{figure}


\begin{thebibliography}{**}
\bibitem{1}
T. Nakano {\it et al.}, Phys. Rev. Lett. {\bf 91}, 012002 (2003);
arXiv:hep-ex/0301020.
\bibitem{2}
DIANA Coll., V.V.~Barmin {\it et al.}, arXiv:hep-exp/0304040 (2003).
\bibitem{3}
CLAS Coll., S.~Stepanyan {\it et al.}, arXiv:hep-ex/0307018 (2003).
\bibitem{4}
SAPHIR Coll., J.~Barth {\it et al.}, arXiv:hep-ex/0307083 (2003).
\bibitem{5}
A.~E.~Asratyan {\it et al.}, arXiv:hep-ex/0309042.
\bibitem{6}
CLAS Coll., V.~Kubarovsky {\it et al.}, arXiv:hep-ex/0311046.
\bibitem{7}
HERMES Coll., A.~Airapetian {\it et al.}, DESY 03-213, December 2003;
arXiv:hep-ex/0312044.
\bibitem{8}
ZEUS Coll., S.~Chekanov,
http://www.desy.de/f/seminar/sem\_schedule.html
\bibitem{9}
SVD Coll., A.~Aleev {\it et al.}, arXiv:hep-ex/0401024.
\bibitem{10}
D.~Diakonov {\it et al.},  Z. Phys. {\bf A 359}, 305 (1997).
\bibitem{11}
T. H. R.~Skyrme,  Nucl. Phys. {\bf 31}, 556 (1962).
\bibitem{12}
A.A.~Arkhipov, arXiv:hep-ph/0309327 (2003); preprint IHEP 2003-37,
Protvino, 2003, available at
http://dbserv.ihep.su/\~{}pubs/prep2003/ps/2003-37.pdf
\bibitem{13}
ZEUS Coll., S.~Chekanov {\it et al.}, arXiv:hep-ex/0403051.
\bibitem{14}
S.~V.~Chekanov (for the ZEUS Collaboration), arXiv:hep-ex/0404007.
\bibitem{15}
P.~Zh.~Aslanyan, V.~N.~Emelyanenko, G.~G.~Rikhkvitzkaya,
arXiv:hep-ex/0403044.
\bibitem{16}
W.~R.~Gibbs, arXiv:nucl-th/0405024.
\end{thebibliography}
\end{document}